\documentclass[11pt]{article}
\usepackage{amsmath,amsthm,latexsym,amssymb,amsfonts,parskip,epsfig}

\usepackage{palatino}
\usepackage[mathcal]{euler}

\newcommand{\Z}{\mathbb{Z}}                  
\newcommand{\N}{\mathbb{N}}                  

\DeclareMathOperator{\res}{Res}

\title{Entropy calculation for a toy black hole}
\author{Hanno Sahlmann, Spinoza Institute and ITP, Utrecht University}
\date{{\small Preprint ITP-UU-07/45, SPIN-07/33}}

\begin{document}
\maketitle
\begin{abstract}
In this note we carry out the counting of states
for a black hole in loop quantum gravity, however
assuming an equidistant area spectrum. We find that
this toy-model is exactly solvable, and we show that
its behavior is very similar to that of the correct
model. Thus this toy-model can be used as a nice and
simplifying `laboratory' for questions about the
full theory.
\end{abstract}
\section{Introduction}
The present paper is concerned with the description of black holes
and calculation of their entropy in the framework of loop quantum
gravity (see \cite{Thiemann:2001yy, Ashtekar:2004eh} for some general
introduction to loop quantum gravity).
The literature on this subject is large and includes (but is
by no means limited to) the pioneering work \cite{Rovelli:1996dv},
the introduction of a precise formalism
\cite{Ashtekar:1997yu,Ashtekar:2000eq}, the reformulation and
approximate solution of the combinatorial problems involved
\cite{Domagala:2004jt,Meissner:2004ju}. Although the basics are by
now quite well understood, there still are surprises in store. One
example are the structures that were found in a computer analysis of
the spectrum of states \cite{Corichi:2006wn,DiazPolo:2007gr}.

The calculation of the entropy of a non-rotating
black hole in loop quantum gravity boils down to a rather complicated combinatorial
problem. It can be treated to a very good approximation in an
asymptotic regime \cite{Domagala:2004jt,Meissner:2004ju}, and
proportionality of entropy to area has been established.

In the present paper we will
develop a model that drastically
simplifies the technical aspects of the entropy
calculation while -- as our results will show --
retaining many of the qualitative features of the
actual situation. In particular, the combinatorial
problem for our model can be solved \textit{exactly},
so any question that one may have about it
can be answered with relative ease.

The simplifying assumption that we will make is
a rather obvious and simple one. One of the hallmarks of loop
quantum gravity is a complicated, non-equidistant
area spectrum. In particular, the area
eigenvalues for a non-rotating isolated horizon
of a black hole are sums of numbers $A_j$,
\label{se_intro}
\begin{equation}
\label{eq_specc} A_j=8\pi\gamma l_P^2
\sqrt{j(j+1)},\qquad j\in \N/2,
\end{equation}
where $\gamma$ is the Barbero-Immirzi parameter, and $l_P$ the Planck-length.
The $A_j$ are obviously not equidistant, however they become approximately
equidistant for large $j$. Our
approximation in this paper consists in using
\begin{equation}
\label{eq_ours} A_j \doteq  8\pi\gamma l_P^2
\left(j+\frac{1}{2}\right)
\end{equation}
instead of \eqref{eq_specc}, i.e. effectively
changing the area operator of the theory. One may
interpret \eqref{eq_ours} as the first two terms in the
series\footnote{Take $\sqrt{j(j+1)+x}$, expand
around $x=1/4$ and evaluate at $x=0$.}
\begin{equation}
\label{eq_perturb} \sqrt{j(j+1)}=j+\frac{1}{2}
-\frac{1}{4 (2 j+1)}-\frac{1}{16 (2
   j+1)^3}+\ldots
\end{equation}
We are certainly not the first use an
approximation like this. For example, it has been
used in \cite{Domagala:2004jt} to give bounds on
the Barbero-Immirzi parameter.

We should also
point out that it has been argued \cite{jurek} that a very similar
equidistant spectrum,
\begin{equation*}
A'_j=8\pi\gamma l_P^2\, j \qquad j\in \N/2,
\end{equation*}
does arise in loop quantum gravity, upon
quantizing the area of a non-rotating black hole following an
alternative route. We will give results on the entropy for
\textit{this} modification of the
area spectrum in the appendix.

The nice thing about the approximation
\eqref{eq_ours} is that it simplifies the
calculation of black hole entropy in the theory
tremendously. We can easily calculate the generating
function corresponding to the combinatorial
problem of enumerating the horizon states.
From the generating function, a lot of information can then be
obtained, as we will demonstrate.

Because of this simplicity our model may be useful, for example to do a
first quick check on some hypothesis, before attempting to check it for the
actual system.
We demonstrate this by studying -- and
ruling out -- some admittedly far fetched proposal about
describing rotating black holes within this
formalism (Section \ref{se_crazy}).


The paper is organized as follows: In the
following section we calculate the generating
function for the problem and derive various
results about the asymptotic growth of the number
of states and thus the entropy.
In Section \ref{se_concl} we will show
that our results parallel those of
\cite{Meissner:2004ju} for the full spectrum, and
discuss some ramifications. In an appendix we
give results for the modified area spectrum $A'_j$.
\section{Counting}
\label{se counting}
In the literature on the subject, slightly different
things have been counted when calculating black hole entropy
in loop quantum gravity \cite{Domagala:2004jt,Khriplovich:2004kx,Tamaki:2005jp,Ghosh:2006ph,Ghosh:2004wq}.
This has to do with the fact that
one has to distinguish between bulk- and
boundary-states\footnote{Barring some sort of holography (for which there is currently little
evidence in LQG), there are infinitely many different bulk states for a black hole of
a given area, so counting those does not even make mathematical sense.} and this
distinction is not entirely trivial. In practice, there arise two different
way to count the entropy, and both lead to the same results
on a qualitative level.

While we do not want to commit ourselves to either of
these ways to count on physical grounds here
(we may have to say more about this elsewhere), we
still restrict to only one way to count in the present article. This is
merely to keep the presentation straightforward. We do not see any
problem to extend our results to the alternative way of counting
\cite{Khriplovich:2004kx,Tamaki:2005jp,Ghosh:2004wq}.

We will follow the definitions of \cite{Ashtekar:2000eq,Domagala:2004jt}:
Let us call the number of surface states with an area smaller or equal  $a$
$N^{\text{true}}_\leq(A)$ (the superscript `true' is meant to indicate
that this is with respect to the actual area spectrum \eqref{eq_specc}).
$N^{\text{true}}_\leq(A)$ can be obtained \cite{Ashtekar:2000eq}
by counting ordered sequences
$(b_i)_i$ of integers $b_i$ modulo $k$ which sum to zero and satisfy
certain additional requirements, namely: There exist sequences
$(m_i)_i$, $m_i\in \Z_*/2$ and $(j_i)_i$, $j_i\in\N_*/2$ such that\footnote{$k\doteq A/4\pi\gamma l_P^2$
has to be integer -- it represents the level of the Chern-Simons theory on the horizon \cite{Ashtekar:2000eq}.}
\begin{equation*}
b_i=-2m_i \mod k, \qquad\text{and}\qquad m_i\in \{-j_i,-j_i+1,\ldots,j_i\}
\end{equation*}
as well as
\begin{equation}
\label{eq_full}
8\pi\gamma l_P^2 \sum_i\sqrt{j_i(j_i+1)}\leq A.
\end{equation}
According to the philosophy
laid out in the introduction, we will just change the area spectrum, and
keep all else unchanged. We will denote by $N_\leq(A)$ the number of sequences
$(b_i)_i$ of integers $b_i$ modulo $k$ which sum to zero and such that there are
$(m_i)_i$,  and $(j_i)_i$ as above, except for that we ask
\begin{equation*}
8\pi\gamma l_P^2 \sum_i \left(j_i+\frac{1}{2}\right)\leq A
\end{equation*}
instead of \eqref{eq_full}. In \cite{Domagala:2004jt} it was
shown that the definition of $N^{\text{true}}_\leq$ is equivalent to a much
simpler one: $N^{\text{true}}_\leq$ is the number of ordered sequences
$(m_i)_i$, $m_i\in \Z_*/2$ such that
\begin{equation*}
\sum_i m_i=0 \qquad \text{and}\qquad
8\pi\gamma l_P^2 \sum_i\sqrt{|m_i|(|m_i|+1)}\leq A.
\end{equation*}
The same arguments can be applied to $N_\leq(A)$. It is easy to see that it is the number
of ordered sequences
$(m_i)_i$, $m_i\in \Z_*/2$ such that
\begin{equation}
\label{eq_bla}
\sum_i m_i=0 \qquad \text{and}\qquad 8\pi\gamma l_P^2 \sum_i \left(|m_i|+\frac{1}{2}\right)\leq A.
\end{equation}
Let us also define $N(A)$, the number of such sequences that satisfy
\eqref{eq_bla} with `$\leq$' replaced by `$=$'.

It was realized in \cite{Domagala:2004jt,Meissner:2004ju} that the counting problem can be simplified
by implementing the two conditions of \eqref{eq_bla} in separate steps. We will follow this
strategy and define
\begin{equation*}
N(a,j)\doteq\left\lvert \left\{(m_1,m_2,\ldots ),\, m_i\in
\Z_*\, : \, \sum_i
m_i=j, \sum_i (|m_i|+1)= a\right\}\right\rvert.
\end{equation*}
Similarly we define
\begin{equation}
\label{eq_sum}
N_\leq(a,j)=\sum_{i=1}^a N(i,j).
\end{equation}
Note that $N_\leq(A)=N_\leq(A/(4\pi\gamma l_P^2),0)$ etc.

A useful way to think about the counting problem for $N(a,j)$ is the
following: $N(a,j)$ is the number ways to move,
in an arbitrary number of steps, on the integer
lattice $\Z$, from the point $0$ to the point
$j$, such that the total length of the path, plus the number of steps, is
$a$.

The numbers $N(a,j)$ obey a recursion relation
similar to the ones given in
\cite{Meissner:2004ju}. It is simple to calculate $N(a,j)$ for low $a$
using a computer. Here are the first few values:
{\tiny
\begin{equation*}
\begin{array}{ccccccccccccccccccccc}
 0 & 0 & 0 & 0 & 0 & 0 & 0 & 0 & 0 & 0 & 1 & 0 & 0 & 0 & 0 & 0 & 0 & 0
   & 0 & 0 & 0 \\
 0 & 0 & 0 & 0 & 0 & 0 & 0 & 0 & 0 & 0 & 0 & 0 & 0 & 0 & 0 & 0 & 0 & 0
   & 0 & 0 & 0 \\
 0 & 0 & 0 & 0 & 0 & 0 & 0 & 0 & 0 & 1 & 0 & 1 & 0 & 0 & 0 & 0 & 0 & 0
   & 0 & 0 & 0 \\
 0 & 0 & 0 & 0 & 0 & 0 & 0 & 0 & 1 & 0 & 0 & 0 & 1 & 0 & 0 & 0 & 0 & 0
   & 0 & 0 & 0 \\
 0 & 0 & 0 & 0 & 0 & 0 & 0 & 1 & 1 & 0 & 2 & 0 & 1 & 1 & 0 & 0 & 0 & 0
   & 0 & 0 & 0 \\
 0 & 0 & 0 & 0 & 0 & 0 & 1 & 2 & 0 & 2 & 0 & 2 & 0 & 2 & 1 & 0 & 0 & 0
   & 0 & 0 & 0 \\
 0 & 0 & 0 & 0 & 0 & 1 & 3 & 1 & 2 & 3 & 2 & 3 & 2 & 1 & 3 & 1 & 0 & 0
   & 0 & 0 & 0 \\
 0 & 0 & 0 & 0 & 1 & 4 & 3 & 2 & 6 & 2 & 6 & 2 & 6 & 2 & 3 & 4 & 1 & 0
   & 0 & 0 & 0 \\
 0 & 0 & 0 & 1 & 5 & 6 & 3 & 9 & 6 & 9 & 8 & 9 & 6 & 9 & 3 & 6 & 5 & 1
   & 0 & 0 & 0 \\
 0 & 0 & 1 & 6 & 10 & 6 & 12 & 14 & 12 & 18 & 12 & 18 & 12 & 14 & 12 &
   6 & 10 & 6 & 1 & 0 & 0 \\
 0 & 1 & 7 & 15 & 12 & 16 & 26 & 20 & 32 & 25 & 34 & 25 & 32 & 20 & 26
   & 16 & 12 & 15 & 7 & 1 & 0 \\
 1 & 8 & 21 & 22 & 23 & 42 & 38 & 50 & 53 & 54 & 58 & 54 & 53 & 50 &
   38 & 42 & 23 & 22 & 21 & 8 & 1
\end{array}
\end{equation*}}
where $j$ runs horizontally from -10 to 10 and $a$ vertically from 0 to 11.

We will now compute the
generating function
\begin{equation*}
G(g,z)\doteq\sum_{a=0}^\infty \sum_{j=-a}^a
N(a,j) g^a z^j.
\end{equation*}
To that end, we refer back to the description of
$N(a,j)$ in terms of paths on $\Z$. Consider
paths with just one step. There is just one such
path from $0$ to $j$ and it has total length
$|j|$. Hence the \textit{one-step generating
function} is
\begin{equation*}
G_1(g,z)=g\sum_{n=1}^{\infty}(gz)^n +
\left(\frac{g}{z}\right)^n=g^2\left(\frac{1}{z-g}+\frac{z}{1-g z}\right).
\end{equation*}
The generating function for paths with $n$ steps
is just $G_1^n$, and thus we get for the
generating function for our problem of interest
(i.e. paths with arbitrary many steps)
\begin{equation*}
G(g,z)=\sum_{n=1}^{\infty} (G_1(g,z))^n
=\frac{g^2 \left(z^2-2 g z+1\right)}{(g+1) \left(2 z
   g^2-\left(z^2+z+1\right) g+z\right)}.
\end{equation*}
Because the $N_\leq(a,j)$ are partial sums of the $N(a,j)$,
the generating function $G_\leq(g,z)$ can be obtained
as \cite{wilf}
\begin{equation}
\label{eq_magic}
G_\leq(g,z)\doteq\sum_{a=0}^\infty \sum_{j=-a}^a
N_\leq(a,j) g^a z^j=\frac{1}{1-g}G(g,z).
\end{equation}
These generating functions contain information
about the counting problem in a very compact and
accessible form. In the following we will extract
some of this information.
\subsection{The asymptotics of $N(a,0)$ and $N_\leq(a,0)$}
The physical states of the black hole horizon
\cite{Ashtekar:1997yu,Domagala:2004jt}
correspond, in our simplified model, to the
states with $j=0$. Therefore the numbers $N(a,0)$
and $N_\leq(a,0)$
are of special interest. We will calculate their
generating functions $G^{(j=0)}(g)$, $G_\leq^{(j=0)}(g)$
and asymptotic behavior.

The generating function $G^{(j=0)}(g)$ is
the coefficient of $z^0$ in $G(g,z)$,
\begin{equation*}
G^{(j=0)}(g)=\frac{1}{2\pi i}\oint_C
\frac{1}{z}G(g,z)\, \text{d}z.
\end{equation*}
where $C$ is a certain contour. Poles of
$G(g,z)/z$ are
\begin{equation*}
z_0=0,\qquad z_{\pm}=-\frac{-2 g^2+g\pm\sqrt{4 g^4-4 g^3+g^2-2 g+1}-1}{2 g}
\end{equation*}
with residues
\begin{align*}
\res_{z_0}(G(g,z)/z)&=-\frac{g}{g+1},\\
\res_{z_\pm}(G(g,z)/z)&=\pm
\frac{(1-g) g}{(g+1) \sqrt{(g-1) (2 g-1) (2 g^2+g+1)}}.
\end{align*}
Choosing the contour $C$ around $z_0$ and $z_+$
gives
\begin{align}
\label{eq_ngen}
G^{(j=0)}(g)&=\frac{(1-g) g}{(g+1) \sqrt{(g-1) (2 g-1) (2 g^2+g+1)}}
-\frac{g}{g+1}\\
&=2 g^4+2 g^6+6 g^7+8 g^8+12 g^9+34 g^{10}+58 g^{11}+\ldots .\nonumber
\end{align}
According to \eqref{eq_magic}, the generating function for
$N_\leq(a,j=0)$ can be obtained as
\begin{align*}
G^{(j=0)}_\leq(g)&=\frac{1}{1-g}G^{(j=0)}(g)\\
&=2 g^4+2 g^5+4 g^6+10 g^7+18 g^8+30 g^9+64 g^{10}+122
   g^{11}+\ldots .
\end{align*}
Let us now look at the asymptotic behavior.
The coefficients in the series expansion of the second term
in \eqref{eq_ngen} are constant, so they do not contribute
at all to asymptotic growth, and we can focus on the first term.
Its singularity at $g= 1/2$
is the one closest to the origin and thus we suspect
that $\ln (N(a,0))\sim \ln(2) a$ to leading order.
Since
the singularities are algebraic, one uses
\textit{Darboux's lemma} (ex. \cite{wilf}) to verify this. It states
that for a function of the form $f(z)=v(z)(1-z)^\beta$
with $\beta \neq \N$ and $v$ analytic in a region
containing the unit disc, expanding $v$ around
$1$ gives an asymptotic expansion of the
function. In particular
\begin{equation*}
[z^n]f(z) = v(1) [z^n] (1-z)^\beta
+O(n^{-\beta-2}) =
v(1)\binom{n-\beta-1}{n}+O(n^{-\beta-2})
\end{equation*}
where we have introduced the notation
$[z^n](\ldots)$ for the coefficient of $z^n$ in
the series expansion around zero.
\begin{figure}
\centerline{\epsfig{file=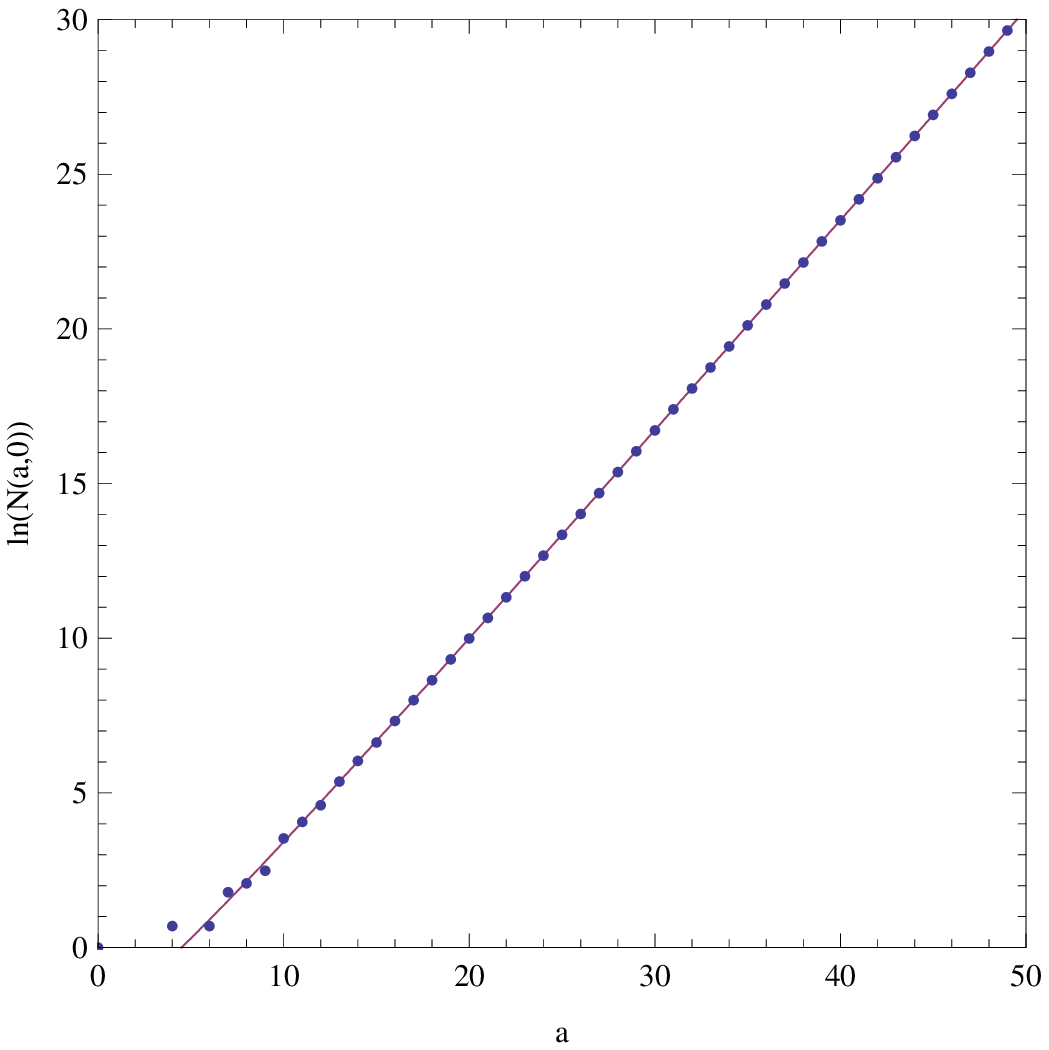, scale=0.6}$\quad$
\epsfig{file=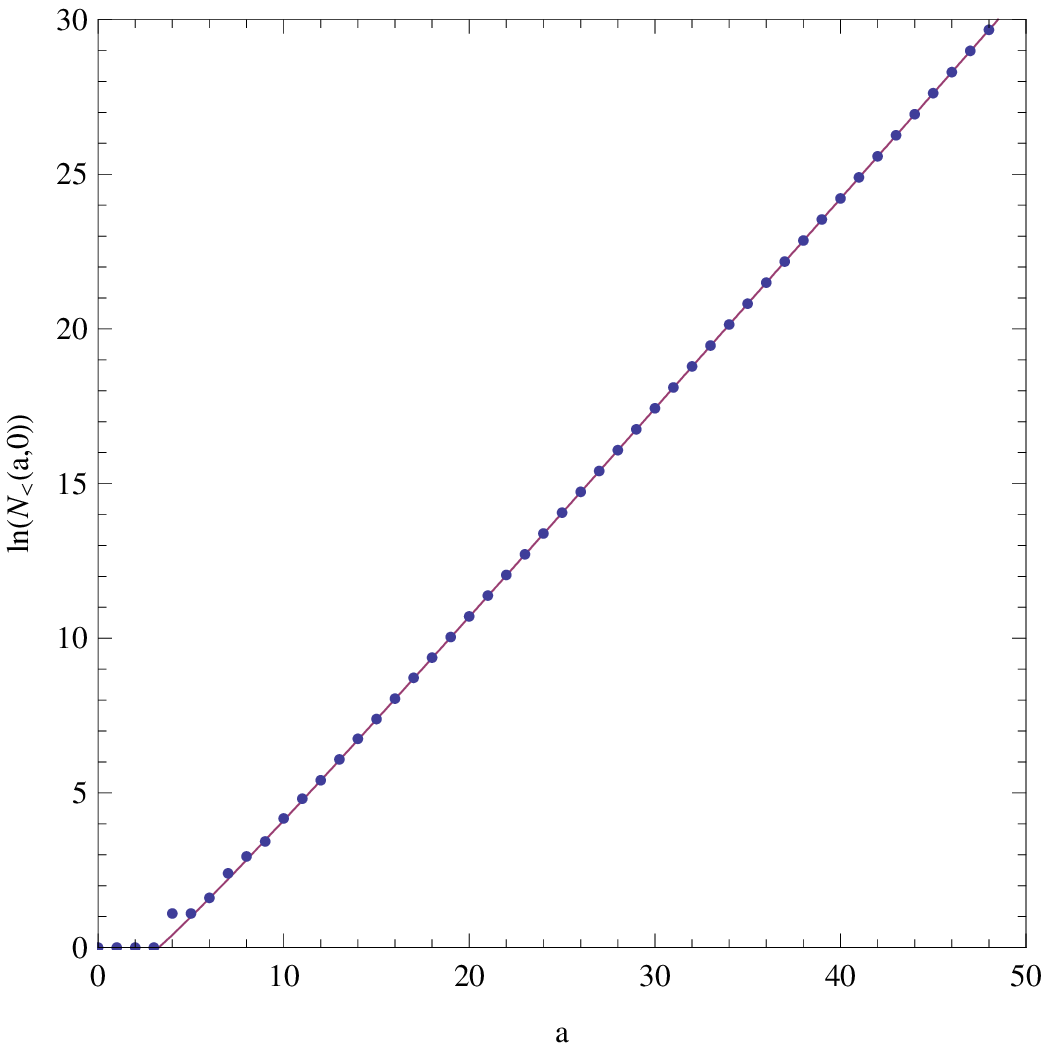, scale=0.6}}
\caption{\label{fi_jzero} $\ln(N(a,0))$ and $\ln(N_\leq(a,0))$ (dots)
and the corresponding asymptotic result (solid line)}
\end{figure}
Applied to our case we find
\begin{align*}
[g^a]G^{(j=0)}(g) &= 2^a [w^a]G^{(j=0)}(w/2)\\
&=2^a [w^a](1-w)^{-\frac{1}{2}}\cdot
\frac{w\sqrt{1-w/2}}{(w+2)\sqrt{1+z/2+z^2/2}}\\
&\sim\frac{1}{6} 2^a \binom{a-1/2}{a}
\sim \frac{1}{6\sqrt{\pi}}\frac{2^a}{\sqrt{a}}
\end{align*}
to highest order. An almost identical calculation gives
\begin{equation*}
[g^a]G^{(j=0)}_\leq(g)\sim \frac{1}{3\sqrt{\pi}}\frac{2^a}{\sqrt{a}}.
\end{equation*}
These approximations are compared
with the actual $N(a,0)$, $N_\leq(a,0)$ in Figure
\ref{fi_jzero}.
\subsection{The asymptotics of unrestricted states}
Our next task is to compute the number of states
without taking into account the restriction
$j=0$,
\begin{equation*}
T(a)\doteq \sum_{j=-a}^{a} N(a,j),\qquad T_\leq(a)\doteq \sum_{j=-a}^{a} N_\leq(a,j).
\end{equation*}
The generating function for $T(a)$ is simply
$G(g,1)$,
\begin{align*}
G(g,1)&=-\frac{2 g^2}{2 g^2+g-1}=2 g^2+2 g^3+6 g^4+10 g^5+\ldots\\
&=\frac{1}{3} \sum_{a=1} (2 (-1)^a+2^a )g^a
\end{align*}
so that we find $T(a)=(2 (-1)^a+2^a )/3$ and hence
and hence
$\ln(T(a))\sim \ln(2) a$.
For $T_\leq(g)$ we find
\begin{align*}
G_\leq(g,1)&=
-\frac{2 g^2}{(1-g) (2 g^2+g-1)}=
2 g^2+4 g^3+10 g^4+20 g^5+\ldots\\
&=\frac{1}{3} \sum_{a=1}(-3+(-1)^a+2^{a+1})g^a
\end{align*}
whence $T_\leq(g)=(-3+(-1)^a+2^{a+1})/3$ and $\ln(T_\leq(a))\sim \ln(2) a$ as well.
\subsection{Asymptotics in both variables}
\label{se_crazy} Finally we take a look at the
joint asymptotics for $a$ and $j$ large and
comparable. This is in part motivated by the
following curious observation: In
\cite{Meissner:2004ju}, the $(a,j)$ asymptotics
is calculated to be
\begin{equation*}
\ln N(a,j) \sim c_1 a+c_2\frac{j^2}{a}
\end{equation*}
where $c_1$ and $c_2$ are certain numerical
constants of order one. This is reminiscent of an
expansion of the Smarr formula for the area of a
Kerr Black hole
\begin{equation*}
A(M,J)=8\pi
M^2\left(1+\sqrt{1-\frac{J^2}{M^4}}\right) =16\pi
M^2-4\pi \frac{J^2}{M^2}+\ldots
\end{equation*}
if one identifies $a$ with $M^2$ and and $j$ with
$J$. $j$ is bounded by $a$ \cite{Domagala:2004jt}
so this identification would not lead out of the
range of allowed spins for a Kerr black hole. It
should also be noted that a similar suggestion
has been before \cite{Krasnov:1998vc}. So, could
it be that the states with $j\neq 0$ describe
rotating black holes?

There are a lot of reasons to doubt that,
including that the meaning of $a$ really should
be area, not mass squared, that states with
$j\neq0$ are un-physical according to the
framework of \cite{Ashtekar:1997yu}, and that
there is a sophisticated treatment of rotating
black holes in loop quantum gravity
\cite{Ashtekar:2004nd} that works quite
differently. In fact we find that the asymptotics we
calculate do not match this hypothesis at all, as follows:

Determining the asymptotics of multivariate
sequences using generating functions is not a
very well known subject and can become quite
technical. There is however a beautifully
developed general theory that one can rely on
(see for example \cite{wilson1} for the generic
case). We will use these results in the form
presented in \cite{wilson2}. The upshot is that
the asymptotics of the multivariate sequence in a
certain direction is governed by one or more
\textit{critical points}, certain singular points
$\underline{z}$ of the generating function. In
the case of a two variable generating function
$G(g,z)=I(g,z)/J(g,z)$, critical points are given
as solutions $\underline{z}=(g_*,z_*)$ to
\begin{equation*}
J(g,z)=0,  \qquad  n_2 g
\frac{\partial}{\partial g} J(g,z) = n_1 z
\frac{\partial}{\partial z} J(g,z)
\end{equation*}
where $(n_1,n_2)\in \Z^2$ gives the direction in
which the asymptotics of the sequence is be
taken. $J$ and $I$ have to satisfy several
properties, for which we refer the reader to
\cite{wilson2}. Which of the critical points
actually contribute to the asymptotics can be
determined by a straightforward but rather
tedious analysis which we will circumvent here.
If a critical point $(g_*,z_*)$ does contribute
to the asymptotics of the sub-sequence $c_n\doteq
c_{n_1 n,n_2n}$ of the multivariate sequence
$c_{m,n}$, it does so by a factor $g^{-n
n_1}_*z_*^{-n_2 n}$ to highest order.

We are interested in the asymptotics of $N(a,j)$
for $\alpha:=j/a$ constant. With the help of
mathematica we compute the critical points for
the problem at hand. There are eight critical
points, all depending $\alpha$. Instead of doing
a lengthy analysis as to which of these solutions
contributes we simply pick the solution with the
right limiting values at $\alpha=0,1$. It reads
\begin{align*}
g_*&=-\frac{(\alpha -1) \left((Y+11) \alpha ^2+(Y+4) \alpha +\left(8 \alpha
   ^2-X (Y-1)\right) \alpha -X Y\right)}{4 \alpha  (\alpha +1)
   \left(-(Y-1) \alpha ^2+X+Y\right)}\\
z_*&=\frac{-(Y-1) \alpha ^2+X+Y}{2 \left(1-\alpha ^2\right)}
\end{align*}
where we have used the abbreviation
\begin{equation*}
X=\sqrt{8 \alpha ^4-11 \alpha ^2+4},\qquad
Y=\sqrt{\frac{\alpha ^2 \left(5 \alpha ^2+2 X-3\right)}{\left(1-\alpha
   ^2\right)^2}}
\end{equation*}
The asymptotic behavior is thus given by
\begin{equation}
\label{eq_asymp} \ln(N(a,\alpha a))\sim
-\ln(g_*) a - \ln(z_*) \alpha a.
\end{equation}
\begin{figure}
\centerline{\epsfig{file=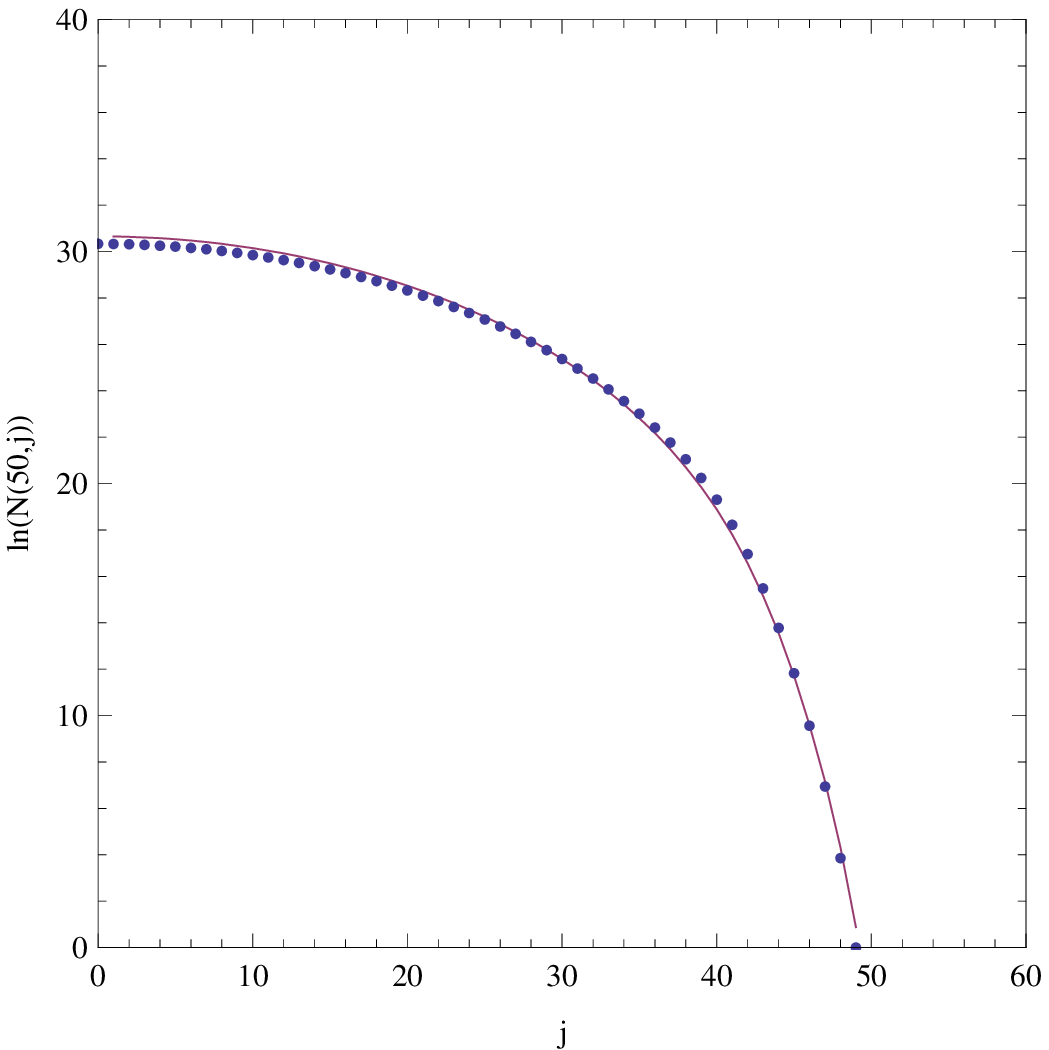,scale=0.6}$\quad$\epsfig{file=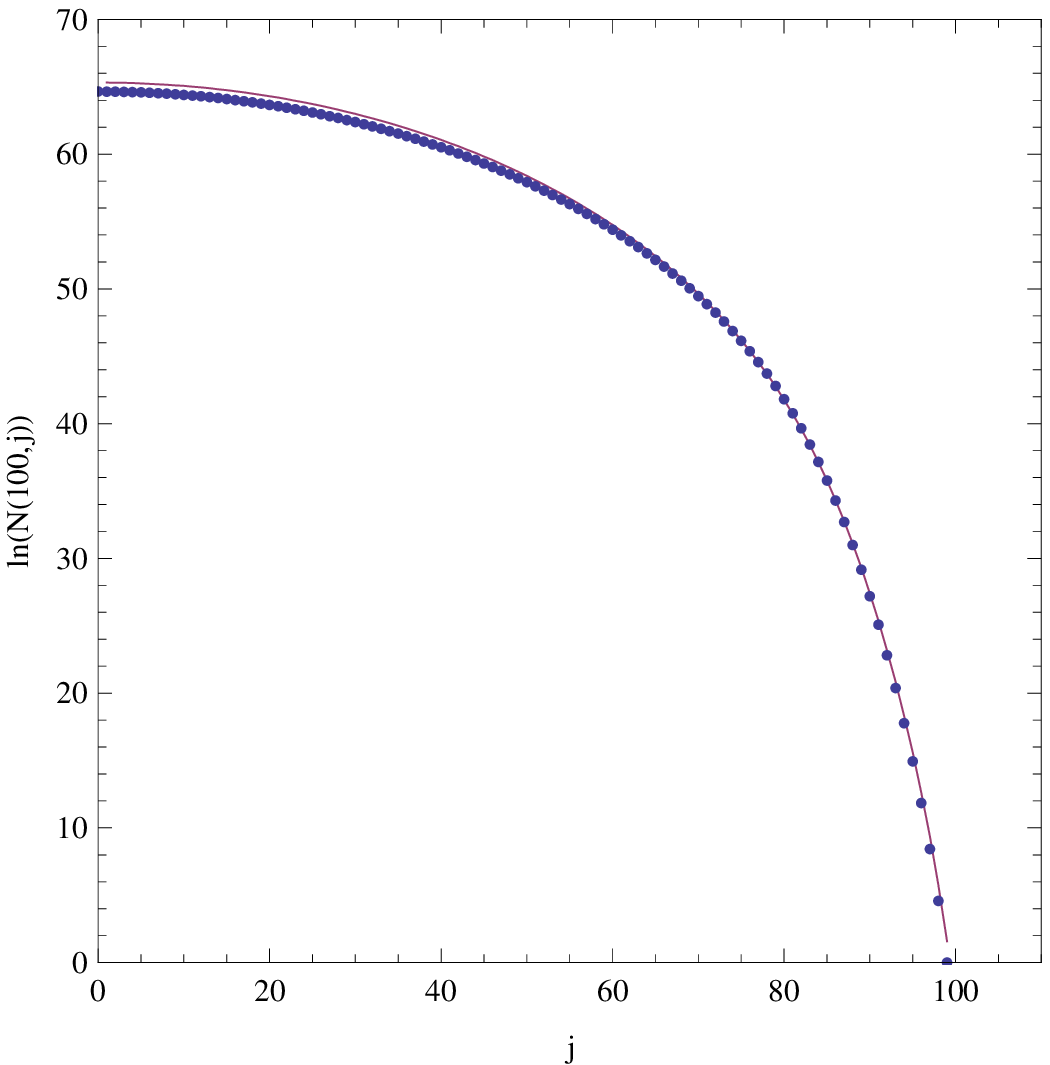,scale=0.6}}
\caption{\label{fi_2dasymp} $\ln(N(a,j))$ (dots)
and the asymptotic result (solid line) compared
for two different values of $a$}
\end{figure}
\begin{figure}
\centerline{\epsfig{file=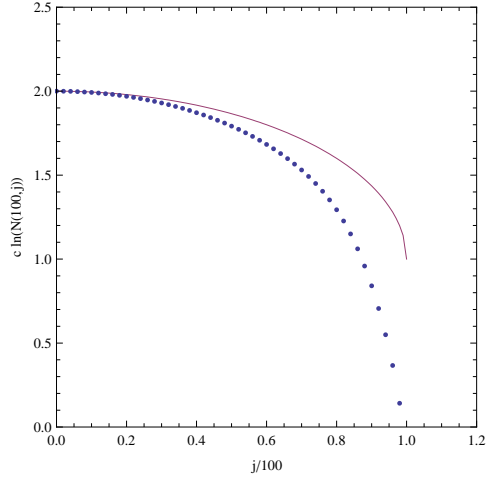,scale=0.6}}
\caption{\label{fi_comp} Comparison of $\ln(N(a,j))$ (dots) and
the function $s$ from \eqref{eq_s} for $a=100$}
\end{figure}
This asymptotic formula works quite well. For
comparison we have plotted \eqref{eq_asymp} and
the actual values for two different values of
$a$.

What about the Smarr formula? We have found that,
to highest order, $ln(N(a,j))$ is indeed a
function of the form $a f(j/a)$. However, the
function $f$ is very complicated and not what one
would expect if the suggested interpretation were
correct. Moreover the numerical fit is quite bad, as Figure
\ref{fi_comp} demonstrates. There we
have plotted the numeric result in terms of
$\alpha=j/a$ (and appropriately normalized),
together with the function
\begin{equation}
\label{eq_s} s(\alpha)=1+\sqrt{1-\alpha^2}.
\end{equation}
Thus we have ample reason to throw out the hypothesis that
we stated in the beginning of this subsection.
\section{Conclusions}
\label{se_concl} Let us summarize our results and
compare them to what is known
\cite{Meissner:2004ju} about the counting for
black holes using the correct area spectrum
\eqref{eq_specc}. To start out, all the
statements made in \cite{Meissner:2004ju} for the
full spectrum are also borne out in our model on
a qualitative level. In particular we see:
\begin{itemize}
\item The number of states grows exponentially with area.
\item The number of (un-physical) states with $j$ arbitrary
grows with same rate as that of the physical
states ($j=0$) to highest order.
\item The highest order growth of $N_\leq(A)$ and
$N(A)$ is the same.
\item The next to leading order term in the logarithm of the number of $j=0$-states
is $-1/2 \ln(a)$.
\end{itemize}
This confirms the present model as a nice and
simplified `laboratory' for questions about the
full theory. Vice versa the present note can be
read as giving confirmation of the results
\cite{Meissner:2004ju}. We should however note
that all the numerical factors appearing in our
model are different from (although close to) the
ones for the correct spectrum. To give an
example, the Barbero-Immirzi-parameter obtained
in \cite{Meissner:2004ju} is $\gamma_M\approx
0.24$ whereas the one that would result if our
model were correct is $\ln(2)/\pi\approx 0.22$.
This is not so surprising if one remembers a
result from \cite{Domagala:2004jt} that shows
that states with $m$ higher than 1/2 certainly
contribute substantially to the overall counting,
however their numbers are increasingly
suppressed with increasing $m$. Thus, among all the states to be
counted, the ones for which our approximation
\eqref{eq_perturb} is relatively bad, are most
numerous.

It is curious that the coefficient of the
$\ln$(area)-term in the entropy seems to be very
robust, as it is -1/2 in the present framework as
well as in
\cite{Meissner:2004ju}, \cite{Corichi:2006wn}, and elsewhere.

There is one last point worth mentioning, one
that on first sight seems trivial: Since the
spectrum is equidistant, the number of black hole
states, and hence the entropy will grow in
discrete steps. The height of the steps may vary
with area, but the size of the steps is always
$4\pi\gamma l_P^2$. Surprisingly, a similar (but certainly
more complex) behavior has been observed
\cite{Corichi:2006wn,DiazPolo:2007gr} for the state counting in
the full theory.
One might at first think that this is not an accident.
After all, the `ladder'
observed in \cite{Corichi:2006wn,DiazPolo:2007gr} may perhaps be
understood as a \textit{perturbation} of the
half integer steps
in the present model, much as the full spectrum
\eqref{eq_specc} can be viewed as a perturbation
\eqref{eq_perturb} around the equidistant one.
We will investigate this point in much more detail in a
forthcoming paper \cite{hs}. The upshot is that the
explanation of the `ladder' is not that simple. It \textit{is}
related to properties of the area spectrum, in particular also
to \eqref{eq_perturb}, but in a rather involved way.
\section*{Acknowledgements}
I want to thank Willem Westra for suggesting the use of
generating functions and for a lot of help in
learning how to use them. I am grateful to Parthasarathi
Mitra for pointing out an error in the bibliography
of an earlier version of this paper.

I also gratefully acknowledge funding for this
work through a Marie Curie Fellowship of the
European Union.
\begin{appendix}
\section{An alternative modified area spectrum}
The form of the modified area spectrum \eqref{eq_ours} in the
main text was chosen so as to approximate the actual
spectrum arising in loop quantum gravity as good
as possible while affording drastic simplification by being equidistant.
There is another
equidistantly spaced modification of the area spectrum
that has been considered in the literature. It differs form
\eqref{eq_ours} by removing the constant $1/2$,
\begin{equation}
\label{eq_jureks}
A_j \doteq  8\pi\gamma l_P^2\, j.
\end{equation}
It can be argued that this spectrum does
arise in loop quantum gravity, upon
quantizing the area of a non-rotating black hole following an
alternative route \cite{jurek}.\footnote{This alternative
quantization is only possible for the black hole horizon.
For other surfaces, only  \eqref{eq_specc} applies.}
In this appendix we will derive results for the
asymptotic behavior of the entropy obtained with
the spectrum \eqref{eq_jureks}. We will however be more
terse than in the main text. We will get rid of all
units and some constants by considering the
quantity
\begin{equation*}
N(a,j)=\left\lvert\left\{(m_1,m_2,\ldots ),\quad m_i\in
\Z\setminus \{0\}\quad : \quad \sum_i
m_i=j, \sum_i |m_i|=a\right\}\right\rvert .
\end{equation*}
A useful way to think about this is the
following: $N(a,j)$ is the number of ways to move,
in an arbitrary number of steps, on the integer
lattice $\Z$, from the point $0$ to the point
$j$, such that the total length of the path is
$a$.

The numbers $N(a,j)$ obey a recursion relation
similar to the ones given in
\cite{Meissner:2004ju}. Up to simple changes of
variables, they are A035002 in \cite{database},
and they are closely related to the combinatorial
problems in \cite{coker}.\footnote{See especially
Section 7 in \cite{coker} where the object under
study is essentially $N(a,0)$.}

It is simple to calculate $N(a,j)$ for low $a$
using a computer. Here are the first few values:
{\footnotesize
\begin{equation*}
\begin{array}{ccccccccccccccc}
 0 & 0 & 0 & 0 & 0 & 0 & 0 & 1 & 0 & 0 & 0 & 0 & 0 & 0 & 0 \\
 0 & 0 & 0 & 0 & 0 & 0 & 1 & 0 & 1 & 0 & 0 & 0 & 0 & 0 & 0 \\
 0 & 0 & 0 & 0 & 0 & 2 & 0 & 2 & 0 & 2 & 0 & 0 & 0 & 0 & 0 \\
 0 & 0 & 0 & 0 & 4 & 0 & 5 & 0 & 5 & 0 & 4 & 0 & 0 & 0 & 0 \\
 0 & 0 & 0 & 8 & 0 & 12 & 0 & 14 & 0 & 12 & 0 & 8 & 0 & 0 & 0 \\
 0 & 0 & 16 & 0 & 28 & 0 & 37 & 0 & 37 & 0 & 28 & 0 & 16 & 0 & 0 \\
 0 & 32 & 0 & 64 & 0 & 94 & 0 & 106 & 0 & 94 & 0 & 64 & 0 & 32 & 0 \\
 64 & 0 & 144 & 0 & 232 & 0 & 289 & 0 & 289 & 0 & 232 & 0 & 144 & 0 & 64
\end{array}
\end{equation*}}
where $j$ runs horizontally and $a$ vertically.

The \textit{one step generating
function} for $N$ is
\begin{equation}
\label{eq_onestep2}
G_1(g,z)=\sum_{n=1}^{\infty}(gz)^n +
\left(\frac{g}{z}\right)^n=-\frac{g}{g-z}-\frac{z
g}{g z-1}.
\end{equation}
The generating function for paths with $n$ steps
is just $G_1^n$, and thus we get for the
generating function for $N$
\begin{equation*}
G(g,z)=\sum_{n=1}^{\infty} (G_1(g,z))^n=-\frac{g
\left(-z^2+2 g z-1\right)}{3 z g^2-2 z^2 g-2
g+z}.
\end{equation*}
\subsection{The asymptotics of $N(a,0)$}
The generating function for $N(a,0)$ is
(formally) the coefficient of $z^0$ in $G(g,z)$,
\begin{equation*}
G^{(j=0)}(g)=\frac{1}{2\pi i}\oint_C
\frac{1}{z}G(g,z)\, \text{d}z.
\end{equation*}
where $C$ is a certain contour. Poles of
$G(g,z)/z$ are
\begin{equation*}
z_{\pm}=\frac{3 g^2\pm \sqrt{9 g^4-10
g^2+1}+1}{4g},\qquad z_0=0
\end{equation*}
with residues
\begin{equation*}
\res_{z_0}(G(g,z)/z)=-\frac{1}{2},\qquad
\res_{z_\pm}(G(g,z)/z)=\pm\frac{1-g^2}{2 \sqrt{9
g^4-10 g^2+1}}
\end{equation*}
Choosing the contour $C$ around $z_0$ and $z_+$
gives
\begin{align*}
G^{(j=0)}(g)&=\frac{1-g^2}{2 \sqrt{9 g^4-10
g^2+1}}-\frac{1}{2}=
-\frac{\sqrt{g^2-1}}{2\sqrt{(g-1/3)(g+1/3)}}-\frac{1}{2}\\
&= 2 g^2+14 g^4+106 g^6+838 g^8+6802
g^{10}+O\left(g^{11}\right)
\end{align*}
The singularities $g=\pm 1/3$ suggest that $\ln
(N(a,0))\propto \ln(3) a$ to leading order.
`Darboux's lemma' confirms this. We find indeed
\begin{align*}
[g^a]G^{(j=0)}(g) &=[v^{a/2}]G^{(j=0)}(\sqrt{v})
=\frac{1}{2} 9^{a/2} [u^{a/2}] (1-u)^{-1/2}\sqrt{9-u}\\
&\propto\sqrt{2} 3^a \binom{a/2-1/2}{a/2} \propto
\frac{2}{\sqrt{\pi}}\frac{1}{\sqrt{a}} 3^a
\end{align*}
to highest order.
\subsection{The asymptotics of unrestricted states}
Our next task is to compute the number of states
without taking into account the restriction
$j=0$,
\begin{equation*}
T(a)\doteq \sum_{j=-a}^{a} N(a,j).
\end{equation*}
The generating function for $T(a)$ is simply
$G(g,1)$,
\begin{equation*}
G(g,1)=\frac{2 g}{1-3 g}=2 \sum_{a=1}^{\infty}
3^{a-1} g^a
\end{equation*}
so that we find $T(a)=2\cdot 3^{a-1}$ and hence
$\ln(T(a))\propto \ln(3) a$.
\subsection{Extremal states}
Out of curiosity, let us also calculate the
number of `extremal configurations', $N(a,a)$. It
this is easiest by going back to
\eqref{eq_onestep2} and modifying it
appropriately. In this case we only need one
variable and we have
\begin{equation*}
G_1(g)=\sum_{n=1}^{\infty}g^n=\frac{g}{1-g}
\end{equation*}
whence
\begin{equation*}
G(g)=\sum_{n=1}^{\infty}\frac{g^n}{(1-g)^n}=\frac{g}{1-2
g}=\sum_{a=1} 2^{a-1}g^a
\end{equation*}
(This is the well known formula for the number of
ordered partitions of an integer $a$.) Hence
$N(a,a)=2^{a-1}$ and $\ln(N(a,a))\propto\ln(2)
a$.
\subsection{Asymptotics in both variables}
Finally we take a look at the
asymptotics of $N(a,j)$
for $\alpha:=j/a$ constant. With the help of
\textit{Mathematica} we compute the critical points for
the problem at hand. There are four critical
points, all depending $\alpha$. Instead of doing
a lengthy analysis as to which of these solutions
contributes we simply pick the solution with the
right limiting values at $\alpha=0,1$. It reads
\begin{equation*}
g_*=\frac{-L+3 \alpha +4}{3 (\alpha
+1)}\frac{\sqrt{1-\alpha ^2}}{\sqrt{-2 \alpha
^2+2 L \alpha +4}},\qquad
z_*=\sqrt{\frac{(L-\alpha ) \alpha +2}{2-2 \alpha
^2}}
\end{equation*}
where we have used the abbreviation
\begin{equation*}
L=\sqrt{4-3 \alpha ^2}
\end{equation*}
The asymptotic behavior is thus given by
\begin{equation*}
\ln(N(a,\alpha a))\propto
-\ln(g_*) a - \ln(z_*) \alpha a.
\end{equation*}
\begin{figure}
\centerline{\epsfig{file=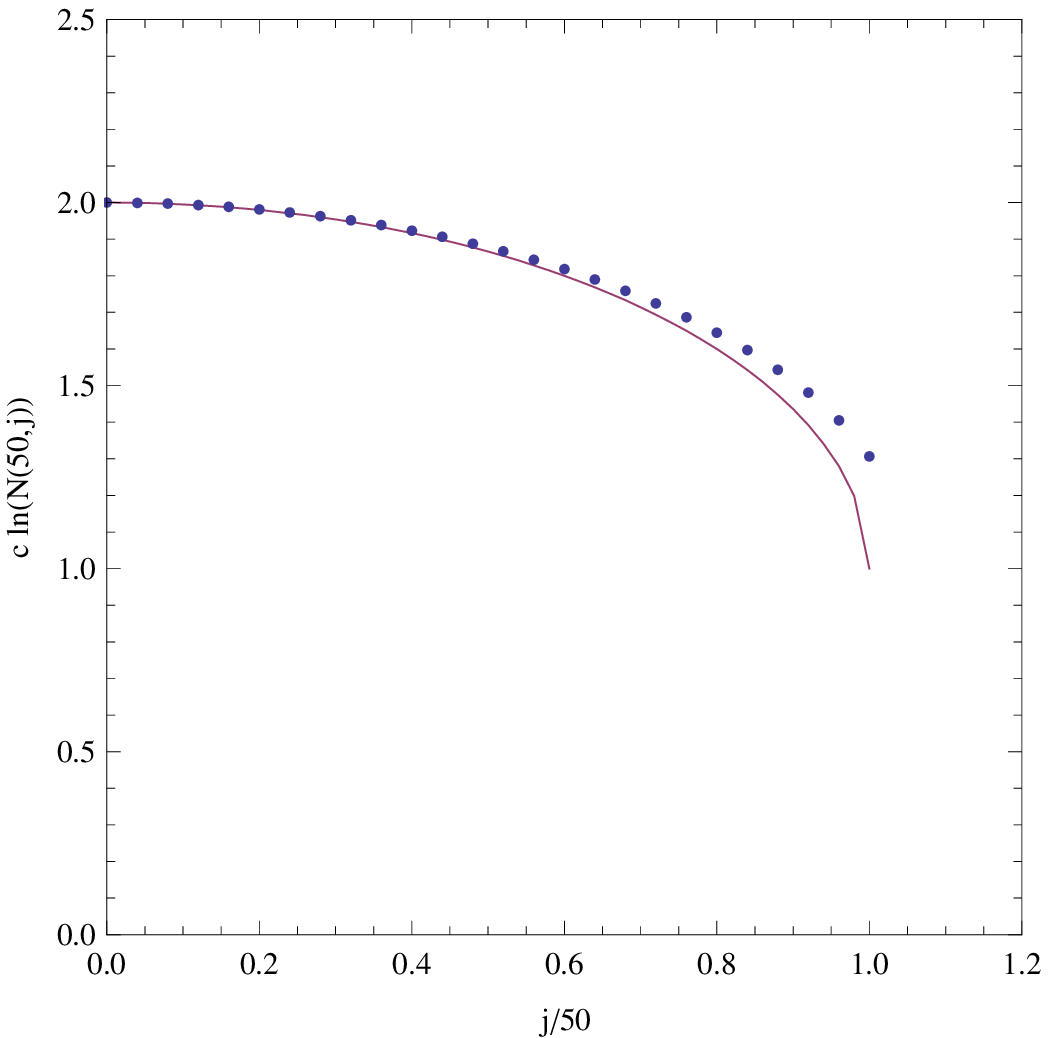,scale=0.6}$\quad$\epsfig{file=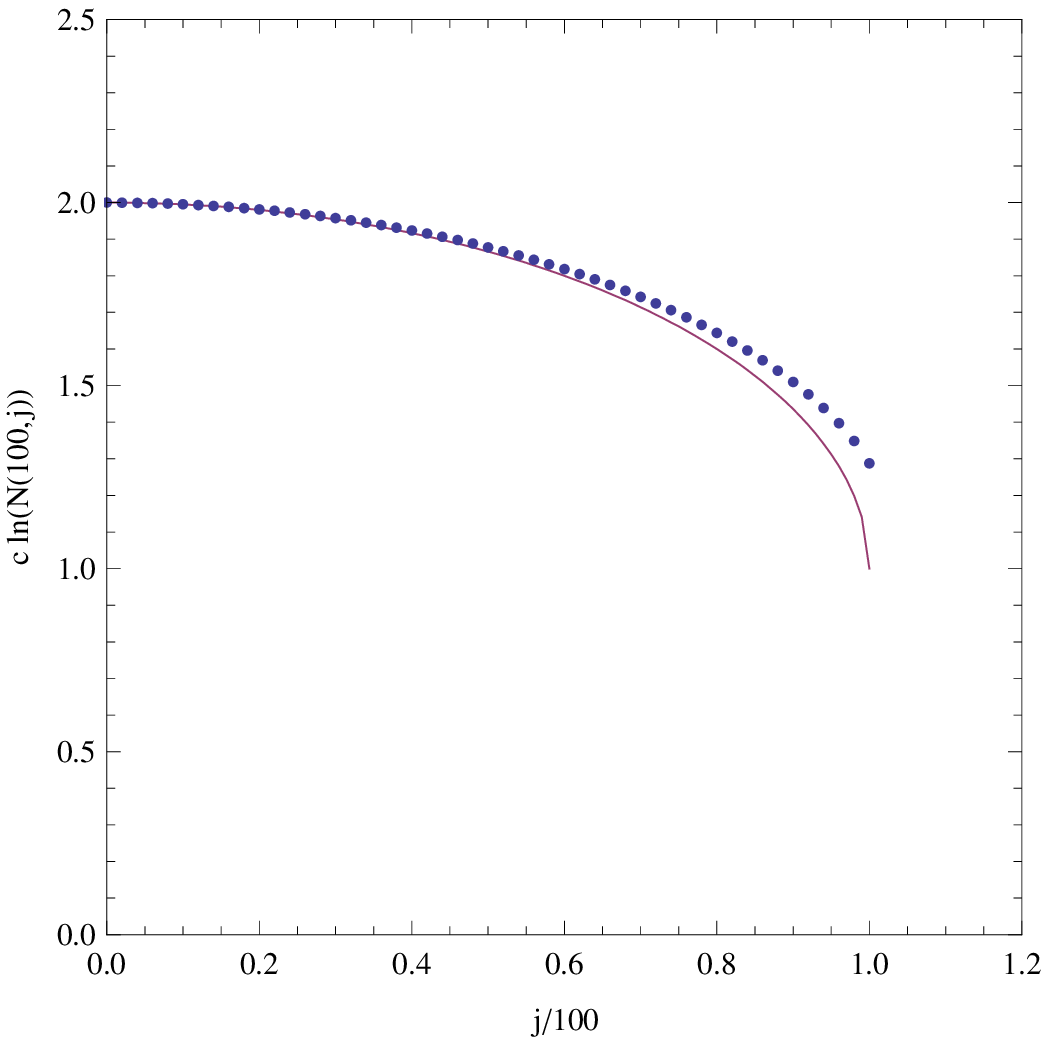,scale=0.6}}
\caption{\label{fi_comp2} $\ln(N(a,j))$ (dots) and
the function $s$ from \eqref{eq_s2} compared for
two different values of $a$}
\end{figure}
What about the Smarr formula in this case? We have found that,
to highest order, $ln(N(a,j))$ is indeed a
function of the form $a f(j/a)$. However, the
function $f$ is very complicated and not what one
would expect if the suggested interpretation were
correct. However, the situation is not
catastrophic either: In Figure \ref{fi_comp2} we
have plotted the numeric result in terms of
$\alpha=j/a$ (and appropriately normalized),
together with the function
\begin{equation}
\label{eq_s2} s(\alpha)=1+\sqrt{1-\alpha^2}.
\end{equation}
One sees that the numerical correspondence is not
bad, but certainly not convincing.
\end{appendix}


\end{document}